\documentclass[twocolumn,showpacs,preprintnumbers,amsmath,amssymb]{revtex4}

\usepackage{graphicx}
\usepackage{dcolumn}
\usepackage{bm}
\usepackage{slashed}

\usepackage{axodraw}
\usepackage{color}

\begin{document}

\title{Thermal Behaviour of $\pi-\pi$ Scattering Lengths in the Nambu--Jona-Lasinio model}

\author{M. Loewe}
 \email{mloewe@fis.puc.cl}

\author{Jorge Ruiz A.}%
 \email{jlruiza@uc.cl}
\affiliation{Facultad de F\'\i sica, Pontificia Universidad
Cat\'olica de Chile,\\ Casilla 306, Santiago 22, Chile. }

\author{J. C. Rojas.}%
 \email{jurojas@ucn.cl}
\affiliation{Departamento de F\'\i sica, Universidad
Cat\'olica del Norte,\\ Angamos 0610, Antofagasta, Chile. }


\date{\today}

\begin{abstract}

We calculate the thermal evolution of $\pi-\pi$ scattering lengths, in the frame of the Nambu--Jona-Lasinio model. The thermal corrections were calculated at the one loop level using Thermofield Dynamics. We present also results for the pion thermal mass. Our procedure implies the modeling of a propagating scalar meson as a resumation of chains of quark bubbles, which is presented explicitly. We compare our results with previous analysis of this problem in the frame of different theoretical approaches.

\end{abstract}

\maketitle

\section{Introduction}

The Nambu--Jona-Lasinio model (NJL) is an effective, non renormalizable, low energy description of hadron dynamics
(\cite{NJL1}, \cite{Klevansky}). The main motivation for this model is the natural way in which the
phenomenon of chiral symmetry breaking appears. The model has the same global symmetries of QCD,
being therefore appropriate for describing low energy hadronic phenomenology. Of special interest are processes
involving pions, since they play a quite special role among hadrons. They are the lightest hadrons, being also
pseudoscalar Goldstone bosons.

In heavy ion relativistic collisions, it turns out that the pion
multiplicity dominates by far any other hadronic signature, being
dominant in the particle content of the central rapidity region,
where we expect the quark gluon plasma to be created. This
suggests that pion interactions, i.e. scattering processes, will
be important. Since those scattering processes occur in the
presence of a thermal environment, our discussion should be
presented in the frame of finite temperature quantum field theory.
The main parameters associated to low energy $\pi$-$\pi$
scattering are the so-called scattering lengths, which depend on
the isospin channel. The first measurements of $\pi$-$\pi$
scattering lengths were done by Rosselet et al \cite{Rosselet}. A
review about the present experimental status of $\pi$-$\pi$
scattering lengths can be found in \cite{Urets}.

In this paper we address once again the problem of determining the thermal
behavior of the $\pi$-$\pi$ scattering lengths, in the frame of the NJL model, using the Thermofield Dynamics (TFD) formalism. For details about this formalism see \cite{bellacdas}, \cite{FG1}. The novelty of our discussion is mainly technical, including the resumation in TFD of bubble diagrams associated to a propagating sigma meson, how to regularize products of distributions, as well as other non trivial intermediate steps that appear in the calculation of the diagrams.

It should be noticed that the thermal behavior of the $\pi$-$\pi$
scattering lengths in the NJL model has being discussed long time
ago in the imaginary time formalism, or Matsubara approach
\cite{Quack}. Our results, specially for $a_2(T)$, do not agree
with this paper. We do agree qualitatively, however, with the
prediction for the thermal behaviour of $\pi$-$\pi$ scattering
lengths in the frame of Chiral Perturbation Theory \cite{Kaiser},
and with a recent analysis based on the Linear Sigma Model
\cite{LM}. Our results indicate that the variation of the
scattering lengths as function of temperature, in particular for
$a_2(T)$, begins at much lower values of temperature than the
behavior found in \cite{Kaiser} and \cite{LM}.

The plan of this article is the following: after presenting the general formalism, we classify the relevant diagrams for getting the $\pi$-$\pi$ scattering lengths according to the NJL model. Then, we proceed to extend our calculation to the finite temperature scenario. For this purpose, a resumation procedure of quark bubbles is presented in order to describe the thermal sigma meson. As a byproduct of our analysis we found also the thermal evolution of the pion mass. Finally, we compare our results with previous articles about this subject.


\section{Scattering lengths and the relevant diagrams at zero temperature}

In general, the $\pi$-$\pi$ scattering amplitude can be parametrized according to

\begin{align}
T_{\alpha\beta;\delta\gamma}&=A(s,t,u)\delta_{\alpha\beta}\delta_{\delta\gamma}+A(t,s,u)\delta_{\alpha\gamma}\delta_{\beta\delta}\nonumber\\
&+A(u,t,s)\delta_{\alpha\delta}\delta_{\beta\gamma}.
\end{align}

\noindent where the $\alpha$, $\beta$, $\gamma$, $\delta$ denote
isospin components.

By using appropriate projection operators it is possible to get
the following isospin dependent scattering amplitudes



\begin{align}
T^{0}&=3A(s,t,u)+A(t,s,u)+A(u,t,s)\\
T^{1}&=A(t,s,u)-A(u,t,s)\\
T^{2}&=A(t,s,u)+A(u,t,s),
\end{align}
where $T^I$ denotes a scattering amplitude in a given isospin channel.\\

Below the inelastic threshold the partial scattering amplitudes
can be parametrized as \cite{Gasser}.


\begin{equation}
T_{\ell}^{I}=\left(\frac{s}{s-4m\pi^2}\right)^{\frac{1}{2}}\frac{1}{2i}\left(e^{2i\delta_{\ell}^{I}(s)}-1\right),
\end{equation}

\noindent where $\delta_{\ell}$ is a phase-shift in the $\ell$
channel. In fact our last
expression can be expanded according to

\begin{equation}
\Re\left(T_{\ell}^{I}\right)=\left(\frac{p^{2}}{m_{\pi}^{2}}\right)^{\ell}\left(a_{\ell}^{I}+\frac{p^2}{m_{\pi}^{2}}b_{\ell}^{I}+...\right).
\end{equation}

The parameters $a_{\ell}^{I}$ and $b_{\ell}^{I}$ are the
scattering lengths and scattering slopes, respectively. In general
the scattering lengths satisfy $|a_{0}|>|a_{1}|>|a_{2}|>...$. If we
are only interested in the scattering lengths $a_0^I$, it is
enough to calculate the scattering amplitude $T^I$ in the static
limit or at threshold i.e. when $s \to 4m_\pi^2$, $t\to 0$ and $u\to
0$

\begin{equation}
a_{0}^{I}=\frac{1}{32\pi}T^{I}\left(s \to 4m_{\pi}^2,t\to 0, u\to0\right).
\end{equation}

\begin{figure}
\centering
\includegraphics[angle=0, width=0.47\textwidth]{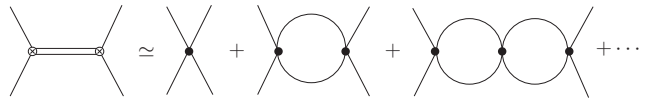}
\caption{Effective meson exchange modeled as a chain of quark bubbles.}
\label{fig:sumageom}
\end{figure}

Our Lagrangian is given by
\begin{equation}
\mathcal{L}_{NJL}=\overline{\psi }\left( i
\slashed{\partial}
-m_{0}\right) \psi +G\left( \left( \overline{\psi }\psi \right) ^{2}+\left(
\overline{\psi }i\gamma _{5}\mathbf{\tau }\psi \right) ^{2}\right),
\end{equation}
where $\psi$ denotes the quark fields, $G$ is a coupling constant and $m_0$ denotes the current quark mass. We will also consider an effective coupling $g_{\pi qq}$ between the internal quark lines and the external pions. Notice that the model does not include a field associated to the scalar sigma meson. The propagator of the sigma meson is represented in the random phase approximation (RPA) trough a geometrical sum of quark bubbles's chains. See fig. \ref{fig:sumageom}, where we show the effective exchange of a sigma meson in the $s-$channel.

\begin{figure}
\centering
\includegraphics[angle=0, width=0.47\textwidth]{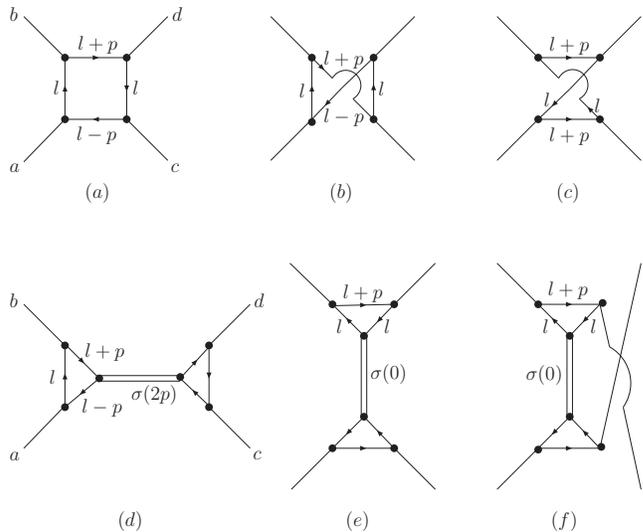}
\caption{Relevant Feynman diagrams for the calculation of the $\pi-\pi$ scattering lengths.}
\label{fig:scattpp}
\end{figure}

The lowest order diagrams that contribute to the $\pi$-$\pi$ scattering lengths in the NJL model are shown in fig. \ref{fig:scattpp}, where the lines with an arrow inside the loops denote the quarks, and the double lines the sigma meson. The external legs are of course the physical pions with momentum $p$ which couple to the quarks trough the effective coupling $g_{\pi qq}$.

At zero temperature these diagrams have been calculated by Schulze \cite{Schulze}.

It is important to remark that the fermion propagators depend on
the constituent quark mass $m$, instead of the current quark mass
$m_0$, because our calculations are done in the chiral broken
phase. The quarks acquire their constituent mass due to a
condensation of quark-antiquark pairs that appear in a completely
analogous way to the Cooper pairs in the BCS theory of
superconductivity. This quark-antiquark pairing mechanism implies
the appearance of the constituent mass trough the so called gap
equation
\begin{equation}
1=\frac{m_{0}}{m}+8iGN_{c}N_{f}\int \frac{d^{4}l}{\left( 2\pi \right) ^{4}}%
\frac{1}{l^{2}-m^{2}+i\varepsilon },  \label{eq:gap}
\end{equation}
where $N=N_cN_f$, being $N_c=3$ the numbers of colors and $N_f$ the numbers of flavors. In our case we will take $N_f=2$.

\section{Thermofield Dynamics and the effective sigma meson propagator.}

In TFD we have to double the number of degrees of freedom in order to express any thermal average in the ensemble as a vacuum expectation value in a populated vacuum. The extra fields are called thermal ghosts and, therefore, the propagators in general will be given by a two by two matrix. In the case of fermions, the matrix propagator has the form
\begin{equation}
S_{F}\left(l\right)=
\begin{pmatrix}
\displaystyle S_{11} & S_{12} \\
S_{21} & S_{22}
\end{pmatrix},  \label{eq:propTFD}
\end{equation}
where
\begin{eqnarray}
S_{11}&=&\left(\slashed{l}+m\right)\left(\displaystyle \frac{i}{l^{2}-m^{2}+i\varepsilon }-\frac{2\pi \delta \left(
l^{2}-m^{2}\right) }{e^{|l_{o}|/T}+1}\right) \notag \\
S_{12}&=&\left(\slashed{l}+m\right)\displaystyle \frac{\varepsilon \left( l_{0}\right) 2\pi
\delta \left( l^{2}-m^{2}\right) e^{|l_{o}|/2T}}{e^{|l_{o}|/T}+1}=S_{21} \notag \\
S_{22}&=&\left(\slashed{l}+m\right)\left(\displaystyle \frac{i}{l^{2}-m^{2}-i\varepsilon }+\frac{2\pi \delta \left(
l^{2}-m^{2}\right) }{e^{|l_{o}|/T}+1}\right), \notag\\
\end{eqnarray}
where $\varepsilon$ is the sign function. Note that the $S_{11}$
element is exactly the fermion Dolan--Jackiw propagator \cite{DJ}.
We may define $S_{11}(p)=S_{F}^0(p)+S_{F}^\beta(p)$, separating
explicitly the zero temperature from the finite temperature
contribution. The other components of the matrix propagator appear
only in internal lines which do not couple to external physical
particles. Therefore, in our case, we will have to deal with the
full matrix propagator structure, at the one loop level, only when
diagrams associated to the exchange of a sigma meson are
considered. As we said, it is given as a sum of chains of quark
bubbles. Notice also that in general, there are two different
coupling constants for the four quark vertices, depending on the
type of quark lines that come together \cite{bellacdas},
\cite{FG1}.

We will now construct the effective thermal sigma meson propagator. In general, the chain of bubbles will imply a resumation of the type $\sum_{n=0}^{\infty }\mathbf{D}^{n}$, where
\begin{equation}
\mathbf{D}=\left(
\begin{array}{cc}
D_{11} & D_{12} \\
D_{21} & D_{22}%
\end{array}%
\right)
\end{equation}%
is a matrix whose elements $D_{ij}$ represent a bubble with an ``open vertex'' of type $i$ at the left side of the bubble, and with a vertex of type $j$ at the right side ($i,j=1,2$). In order to avoid double counting of vertices when powers of $\mathbf{D}$ are considered, we have to introduce quark bubbles with an open vertex only at the left side. It is clear that each element $D_{ij}^{2}$ involves the sum of the two possibilities to build a chain with two bubbles, in such a way that we have an open vertex of type $i$ at the left side and a vertex of type $j$ at the right side. This means that the two possible values for the vertex in between are included. The same happens for every matrix element of some power $n$ of $\mathbf{D}$.

It is convenient to define the following matrix
\begin{eqnarray}
\mathbf{Z}&\equiv& \frac{1}{1-D_{11}-D_{22}+D_{11}D_{22}-D_{12}D_{21}}\nonumber\\
&&\times\left(
\begin{array}{cc}
v_{1}(1-D_{22}) & v_{1}D_{12} \\
v_{2}D_{21} & v_{2}(1-D_{11})%
\end{array}%
\right), \label{eq:Z}
\end{eqnarray}
that essentially corresponds to $\sum_{n=0}^{\infty }\mathbf{D}^{n}$, where, we have explicitly introduced the vertices $v_i$ associated to the left side of the sum of the quark bubbles. A similar treatment was introduced previously in the discussion of thermal renormalons in the $\lambda\phi^4$ theory \cite{FLRV}.

\begin{figure}
\centering
\includegraphics[angle=0, width=0.3\textwidth]{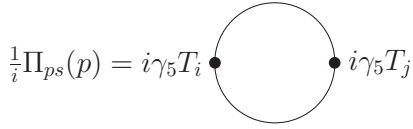}
\caption{The single quark bubble that model a pion in the NJL approach.}
\label{fig:burbuja}
\end{figure}

In order to appreciate the role of the $Z$ matrix let us consider diagrams (d), (e) and (f) of fig. \ref{fig:scattpp}. There we have quark triangles that couple to the effective sigma meson trough open vertices. To be more explicit, let us denote by $\Gamma_i$ the triangle diagram with an internal open vertex of type $i$. Then we introduce the vectors
\begin{equation}
\mathbf{Y}^{\top }=\left(
\begin{array}{cc}
\Gamma _{1} & \Gamma _{2}%
\end{array}%
\right) ,
\qquad
\mathbf{Y}=\left(
\begin{array}{c}
\Gamma _{1} \\
\Gamma _{2}%
\end{array}%
\right).
\end{equation}%
In this way, the complete amplitude for these kind of diagrams, including both the zero as well as the finite temperature parts, is given in this matrix language by
\begin{equation}
\mathcal{T}=\mathbf{Y}^{\top }\mathbf{Z}\mathbf{Y}.
\end{equation}%
In the case of the triangle diagrams $(d), (e)$ and $(f)$ of fig. \ref{fig:scattpp}, we need the following elements
\begin{eqnarray}
D_{11} &=&\left(\frac{1}{i}\Pi^{0}+2\frac{1}{i}\Pi^{\beta }\right)v_{1} \\
D_{12} &=&0 \\
D_{21} &=&0 \\
D_{22} &=&\left(\frac{1}{i}\Pi^{0}-2\frac{1}{i}\Pi^{\beta }\right)v_{2}
\end{eqnarray}%
to build the $\mathbf{Z}$ matrix. In the above expression $-i\Pi^{0} (-i\Pi^{\beta})$ denotes a zero (finite) temperature single bubble contribution. Notice that $D_{12}=D_{21}=0$ because there is no support for the corresponding integrals, as can be checked, since we are using constituent quark masses $m$, being $m>2m_\pi$. In order to handle other singular contributions, like products of delta functions with the same argument, it was necessary to take different external pion momenta, i.e. a kinematical configuration away from the threshold. In this way, singularities disappear and we are able to compute the corresponding integrals, finding that they vanish identically when the limit to the threshold configuration is taken. Because of the same reasons, the triangle diagram $\Gamma_2$ which couples to the sigma meson trough a vertex $v_2$, vanishes. In this way, for the triangle diagrams we found
\begin{eqnarray}
\Gamma _{1} &=&\Gamma _{d}^{0}+4\Gamma _{d}^{\beta } \\
\Gamma _{2} &=&0,
\end{eqnarray}
where
\begin{equation}
\Gamma _{d}^{\beta } =
\frac{2m}{\pi ^{2}}N\left( g_{\pi qq}^{\beta }\right)
^{2}\int_{0}^{\Lambda }dl\frac{l^{2}n_F(E_l)}{E_l(4E_l^2-m_{\pi }^{2})}. \label{eq:tdb}
\end{equation}
where $E_{l}=\sqrt{\mathbf{l}^2+m^2}$ and $n_F$ is the usual Fermi--Dirac distribution
\begin{equation}
n_{F}(z)\equiv\frac{1}{e^{|z_0|/T}+1}.
\end{equation}
We have introduced a temperature dependent coupling $g_{\pi qq}^{\beta }$ between pions and quarks. This coupling can be obtained as
\begin{equation}
\left( g_{\pi qq}^{\beta }\right) ^{-2}=\left( \frac{\partial \Pi _{ps}}{%
\partial p^{2}}+2\frac{\partial \Pi _{ps}^{\beta }}{\partial p^{2}}\right)
\Bigg{|}_{p^{2}=\left( m_{\pi }^{\beta }\right) ^{2}},  \label{eq:gpiqqt}
\end{equation}
where $-i\Pi _{ps}$ is the single quark bubble (pseudoscalar case) shown in fig. \ref{fig:burbuja}. At this point it is worthwhile to remember that such a quark bubble is used to model a pion in the NJL approach. $T$ selects the isospin channel: $T_i=T_j=\tau_3$ corresponds to the $\pi^0$ and/or $T_i=\tau^{(\pm)}$ and $T_j=\tau^{(\mp)}$ are associated to the $\pi^{(\pm)}$. In the previous equation, $m_{\pi }^{\beta }$ is the thermal pion mass which is obtained in the next section.

\section{Temperature corrections to the $\pi$-$\pi$ scattering lengths}
Following the notation introduced by Schulze \cite{Schulze}, all diagrams at zero temperature are given in terms of the three following integrals
\begin{eqnarray}
I\left( p\right) &=&\int^{\Lambda }\frac{d^{4}l}{\left( 2\pi \right) ^{4}}%
\frac{1}{\left( l^{2}-m^{2}+i\varepsilon \right) \left( \left( l+p\right)
^{2}-m^{2}+i\varepsilon \right) }  \nonumber \\
K\left( p\right) &=&\int^{\Lambda }\frac{d^{4}l}{\left( 2\pi \right) ^{4}}%
\frac{1}{\left( l^{2}-m^{2}+i\varepsilon \right) ^{2}\left( \left(
l+p\right) ^{2}-m^{2}+i\varepsilon \right) } \nonumber \\
L\left( p\right) &=&\int^{\Lambda }\frac{d^{4}l}{\left( 2\pi \right) ^{4}}%
\frac{1}{\left( l^{2}-m^{2}+i\varepsilon \right) ^{2}\left( \left(
l+p\right) ^{2}-m^{2}+i\varepsilon \right) ^{2}},  \nonumber\\
\end{eqnarray}%
where $\Lambda$ is a 3-momentum cutoff. After performing the angular and the $l_0$ integrals, we found
\begin{eqnarray}
I\left( p\right)
&=&\frac{i}{2\pi ^{2}}\int_{0}^{\Lambda }dl\frac{l^{2}}{E_{l}\left(
4E_{l}^{2}-p^{2}\right) }\\
K\left( p\right) &=&\frac{i}{8\pi ^{2}}\int_{0}^{\Lambda }dl\frac{%
l^{2}\left( p^{2}-12E_{l}^{2}\right) }{E_{l}^{3}\left(
4E_{l}^{2}-p^{2}\right) ^{2}} \\
L\left( p\right) &=&\frac{i}{4\pi ^{2}}\int_{0}^{\Lambda }dl\frac{%
l^{2}\left( 20E_{l}^{2}-p^{2}\right) }{E_{l}^{3}\left(
4E_{l}^{2}-p^{2}\right) ^{3}}.
\end{eqnarray}

In the NJL approach, the gap equation plays a fundamental role. At finite temperature \cite{Klevansky} it acquires the form
\begin{eqnarray}
1 &=& \frac{m_{0}}{m}+8iGN\int^{\Lambda }\frac{d^{4}l}{\left( 2\pi
\right) ^{4}}\frac{1}{l^{2}-m^{2}+i\varepsilon } \notag\\
&& +8iGN\int^{\Lambda }\frac{d^{3}l%
}{\left( 2\pi \right) ^{3}}\frac{in_{F}\left( E_{l}\right) }{E_{l}}.
\label{eq:gapt}
\end{eqnarray}

\begin{figure}
\centering
\includegraphics[angle=0, width=0.5\textwidth]{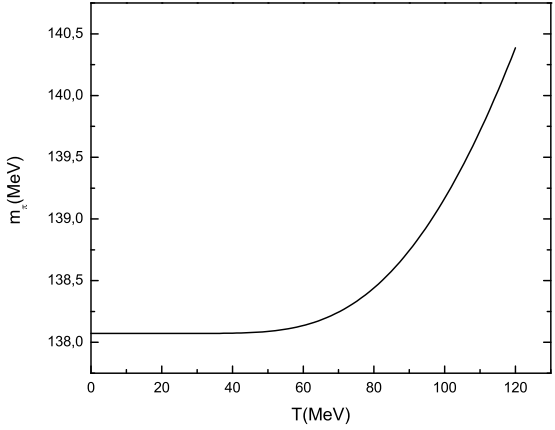}
\caption{Thermal evolution of the pion mass.}
\label{fig:mpit}
\end{figure}

The thermal corrections to the effective sigma meson propagator, at the level of a single quark bubble are shown in the fig. \ref{fig:burbuja}, but where the $i\gamma_5 T$ has been replaced by the identity matrix according to the NJL Lagrangian. This loop is given by
\begin{equation}
\frac{1}{i}\Pi^{\beta }\left( p\right) =2iN\left( \int^{\Lambda }\frac{%
d^{3}l}{\left( 2\pi \right) ^{3}}\frac{n_{F}\left( E_{l}\right) }{E_{l}}%
+\left( 4m^{2}-p^{2}\right) J\left( p\right) \right),
\end{equation}
where $J(p)$ is
\begin{equation}
J\left( p\right) =\frac{1}{2\pi ^{2}}\int_{0}^{\Lambda }dl\frac{%
l^{2}n_F(E_l) }{E_{l}\left( p^{2}-4E_{l}^{2}\right) }.
\label{eq:j}
\end{equation}
Notice that at this level the Dolan--Jackiw propagators are enough. In particular, the thermal contributions arise from the insertion of only one pure thermal propagator $S_F^\beta$ in the loop. The other propagator has to be the normal one at zero temperature. It is easy to see that two thermal propagators do not contribute. The denominator $D_Z(p)$ that appears in the $\mathbf{Z}$ matrix (\ref{eq:Z}) is given by
\begin{equation}
D_Z(p)=1-2G\Pi _{ps}^{0}\left( p\right) -4G\Pi _{ps}^{\beta }\left( p\right).
\end{equation}
Using the thermal mass gap equation (\ref{eq:gapt}) this expression can be written as
\begin{equation}
D_Z(p)=\frac{m_{0}}{m}+4iGNp^{2}\left( I\left( p\right) +2iJ\left( p\right)
\right). \label{eq:abc}
\end{equation}
Following the same procedure, we could also construct an effective pion propagator, which is not necessary for our purpose of getting the $\pi$-$\pi$ scattering lengths. From this analysis, however, we get the thermal pion mass determined from the pole of the effective propagator
\begin{equation}
\left( m_{\pi }^{\beta }\right) ^{2}=-\frac{m_{0}}{m}\frac{1}{4iGN\left(
I\left( m_{\pi }\right) +2iJ\left( m_{\pi }\right) \right) }.
\label{eq:mpit}
\end{equation}
The behavior of $m_{\pi }^{\beta }$ is shown in fig. \ref{fig:mpit}. This allows us to express the quotient $m_0/m$ in terms of the other quantities that appear in the previous equations. Replacing $m_0/m$ in (\ref{eq:abc}) we get
\begin{eqnarray}
D_Z(p) &=& 4iGN \left( p^{2}-4m^{2}\right) \left( I\left( p\right) +2iJ\left(
p\right) \right) \notag\\
&& -4iGN m_{\pi }^{2}\left( I\left( m_{\pi }\right) +2iJ\left(
m_{\pi }\right) \right).  \label{eq:densigma}
\end{eqnarray}
An expression for $m_\pi^\beta$ was also found trough the mass gap equation in the Matsubara or imaginary time formalism by previous authors \cite{HK}. We agree qualitatively with them.

It is important to remark that in equation (\ref{eq:densigma}), $m_\pi$ is the usual pion mass at zero temperature because we are implementing a calculation at the one loop level. The inclusion of $m_\pi^\beta$ in the last expression goes beyond this approximation.

Finally, we get the following expression, consistent at the one loop perturbation theory, for the effective sigma meson propagator
\begin{equation}
D_{\sigma }^{0+\beta }\left( p\right)=\frac{1}{2N}\frac{4iGN}{D_Z(p)}.
\label{eq:propefsigmat}
\end{equation}

In addition to the triangle diagram given in eq. (\ref{eq:tdb}), we need, also associated to the diagram \ref{fig:scattpp}(e), the following quantities
\begin{eqnarray}
\Gamma _{e_{-}}^{\beta } &=&
\frac{4m}{\pi ^{2}}N\left( g_{\pi qq}^{\beta }\right)
^{2}\int_{0}^{\Lambda }\frac{l^{2}n_F(E_l)(4E_l^2+m_{\pi }^{2})}{E_l(4E_l^2-m_{\pi }^{2})^{2}}dl \notag\\
\Gamma _{e_{0}}^{\beta } &=&
\frac{8}{\pi ^{2}}N\left( g_{\pi qq}^{\beta }\right)
^{2}\int_{0}^{\Lambda }\frac{\partial }{\partial m^{2}}\left( \frac{ml^{2}%
n_F(E_l)E_l}{(4E_l^2-m_{\pi }^{2})%
}\right) dl  \nonumber \\
&&-\frac{2}{\pi ^{2}m}N\left( g_{\pi qq}^{\beta }\right)
^{2}M_e,
\end{eqnarray}
where
\begin{equation}
M_e=\int_{0}^{\Lambda }\frac{l^{2}n_F(E_l)(m_{\pi }^{2}m^{2}+(4m^{2}+2m_{\pi
}^{2})E_l^2-8E_l^{4})}{E_l(4E_l^2-m_{\pi }^{2})^{2}}dl.
\end{equation}
For the diagrams $(d), (e)$ and $(f)$, fig. \ref{fig:scattpp}, we get
\begin{eqnarray}
\mathcal{T}_{d}^{0+\beta } &=&D_{\sigma }^{0+\beta }\left( 2m_{\pi }\right)
\left( \Gamma _{d}^{0}+4\Gamma _{d}^{\beta }\right) ^{2}  \nonumber \\
\mathcal{T}_{e}^{0+\beta } &=&D_{\sigma }^{0+\beta }\left( 0\right) \left(
\Gamma _{e}^{0}+\Gamma _{e_{-}}^{\beta }+2\Gamma _{e_{0}}^{\beta }\right)
^{2}  \nonumber \\
\mathcal{T}_{f}^{0+\beta } &=&\mathcal{T}_{e}^{0+\beta },
\end{eqnarray}
where \cite{Schulze}
\begin{eqnarray}
\Gamma _{d}^{0} &=& -8NmI(m_\pi)\left(g_{\pi qq}^0\right)^{2} \nonumber \\
\Gamma _{e}^{0} &=& -8Nm\left(I(0)-m_\pi^2K(m_\pi)\right)\left(g_{\pi qq}^0\right)^{2}.
\end{eqnarray}
$g_{\pi qq}^0$ can be read from the first term in eq. \ref{eq:gpiqqt}. In terms of our integrals $I(p)$ and $K(p)$, it can be rewritten as
\begin{equation}
g_{\pi qq}^0=\left(-Ni\left(I(0)+I(m_\pi)-m_\pi^2K(m_\pi)\right)\right)^{-1/2}.
\end{equation}
We find $g_{\pi qq}^0=3.56$.

Concerning the temperature corrections to the box diagrams $(a), (b)$ and $(c)$ in fig. \ref{fig:scattpp}, we only need the Dolan--Jackiw fermion propagators. For the same reasons we mentioned previously, in each box diagram will survive only the insertion of one pure thermal fermion propagator $S^\beta$, which can be any of the four propagators inside the box. Two, three or four thermal insertions vanish identically. The results are
\begin{eqnarray}
\mathcal{T}_{a}^{0+\beta } &=&\mathcal{T}_{a}^{0}+2\mathcal{T}_{a_{\pm
}}^{\beta }+2\mathcal{T}_{a_{0}}^{\beta }  \nonumber \\
\mathcal{T}_{b}^{0+\beta } &=&\mathcal{T}_{a}^{0+\beta }  \nonumber \\
\mathcal{T}_{c}^{0+\beta } &=&\mathcal{T}_{c}^{0}+4\mathcal{T}_{c_{\pm
}}^{\beta },
\end{eqnarray}
where
\begin{eqnarray}
\mathcal{T}_{a_{\pm}}^{\beta } \! &=& \! \frac{8i}{\pi ^{2}}%
N\left( g_{\pi qq}^{\beta }\right) ^{4}\int_{0}^{\Lambda }\frac{l^{2}n_F(E_l)E_l}{(4E_l^2-m_{\pi }^{2})^{2}}%
dl \nonumber\\
\mathcal{T}_{a_{0}}^{\beta }
\! &=& \! -\frac{4i}{\pi ^{2}}N\left( g_{\pi qq}^{\beta }\right)
^{4}\int_{0}^{\Lambda }\frac{\partial }{\partial m^{2}}\left( \frac{l^{2}n_F(E_l)E_l}{(4E_l^2-m_{\pi }^{2})%
}\right) dl  \nonumber \\
&&-\frac{2im_{\pi }^{2}}{\pi ^{2}}N\left( g_{\pi qq}^{\beta }\right)
^{4}\int_{0}^{\Lambda }\frac{l^{2}n_F(E_l)}{E_l(4E_l^2-m_{\pi }^{2})^{2}}dl.\nonumber\\
\mathcal{T}_{c_{\pm }}^{\beta }
\!\! &=& \!\! \frac{4N\left( g_{\pi qq}^{\beta }\right)
^{4}}{i\pi ^{2}}\!\!\int_{0}^{\Lambda }\!\!\frac{\partial }{\partial m^{2}}\!\!\left( \frac{l^{2}n_F(E_l)E_l(4E_l^2+m_{\pi }^{2})}{(4E_l^2-m_{\pi }^{2})^{2}}\right) \! dl  \nonumber \\
&&-\frac{2m_{\pi }^{2}i}{\pi ^{2}}N\left( g_{\pi qq}^{\beta }\right)
^{4}\int_{0}^{\Lambda }\frac{l^{2}n_F(E_l)\left( m_{\pi }^{2}+12E_l^2 \right) }{E_l(4E_l^2-m_{\pi }^{2})^{3}}dl. \notag\\
\end{eqnarray}
$\mathcal{T}_{a}^{0}$ and $\mathcal{T}_{c}^{0}$ were also obtained in \cite{Schulze}, and are given by
\begin{eqnarray}
\mathcal{T}_{a}^{0} &=&  4N\left( g_{\pi qq}^{\beta }\right) ^{4}\left(m_\pi^2K(m_\pi)-I(0)-I(m_\pi)\right) \nonumber\\
\mathcal{T}_{c}^{0} &=&  8N\left( g_{\pi qq}^{\beta }\right) ^{4}\left(2m_\pi^2K(m_\pi)-I(0)-\frac{1}{2}m_\pi^4L(m_\pi)\right). \nonumber\\
\end{eqnarray}

\begin{figure}
\centering
\includegraphics[angle=0, width=0.5\textwidth]{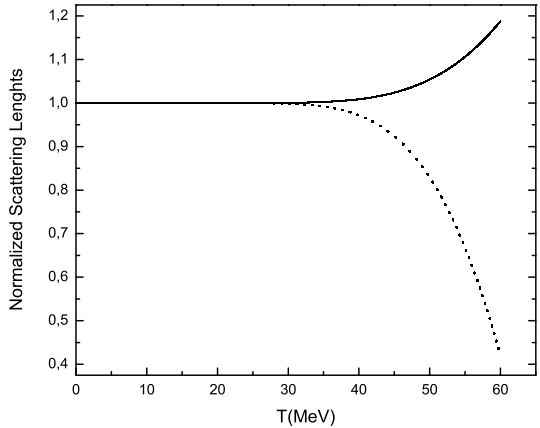}
\caption{Thermal dependence of the $\pi-\pi$ scattering lengths, normalized to its zero temperature value. The upper curve correspond to $a_{0}^{0+\beta}/a_{0}^{0}$, and the lower one to $a_{2}^{0+\beta}/a_{2}^{0}$. Temperatures are in MeV.}
\label{fig:scattlengths}
\end{figure}

Using these results, we are able to find the thermal $\pi-\pi$ scattering lengths. For each diagram in fig \ref{fig:scattpp}, the total
amplitude is given as the sum of the zero temperature and the finite temperature parts. In this way we get
\begin{eqnarray}
\mathcal{T}_{a}^{0+\beta } &=&\mathcal{T}_{a}^{0}+2\mathcal{T}_{a_{\pm
}}^{\beta }+2\mathcal{T}_{a_{0}}^{\beta }  \nonumber \\
\mathcal{T}_{b}^{0+\beta } &=&\mathcal{T}_{a}^{0+\beta }  \nonumber \\
\mathcal{T}_{c}^{0+\beta } &=&\mathcal{T}_{c}^{0}+4\mathcal{T}_{c_{\pm
}}^{\beta }  \nonumber \\
\mathcal{T}_{d}^{0+\beta } &=&D_{\sigma }^{0+\beta }\left( 2m_{\pi }\right)
\left( \Gamma _{d}^{0}+4\Gamma _{d}^{\beta }\right) ^{2}  \nonumber \\
\mathcal{T}_{e}^{0+\beta } &=&D_{\sigma }^{0+\beta }\left( 0\right) \left(
\Gamma _{e}^{0}+\Gamma _{e_{-}}^{\beta }+2\Gamma _{e_{0}}^{\beta }\right)
^{2}  \nonumber \\
\mathcal{T}_{f}^{0+\beta } &=&\mathcal{T}_{e}^{0+\beta },
\end{eqnarray}%
where the superscript 0 refers to zero temperature terms, which are given in \cite{Schulze}. To express them, we have used our functions $I,K$ and $L$. Finally, the scattering lengths are given by
\begin{eqnarray}
a_{0}^{0+\beta } &=&\frac{1}{32\pi }\left( 6\mathcal{T}_{a}^{0+\beta }-%
\mathcal{T}_{c}^{0+\beta }+3\mathcal{T}_{d}^{0+\beta }+2\mathcal{T}%
_{e}^{0+\beta }\right)   \nonumber \\
a_{2}^{0+\beta } &=&\frac{1}{32\pi }\left( 2\mathcal{T}_{c}^{0+\beta }+2%
\mathcal{T}_{e}^{0+\beta }\right).
\end{eqnarray}%
Their thermal dependence is shown in
fig. \ref{fig:scattlengths}.

The parameters we used in the calculation are $\Lambda=631\text{ MeV}$, $m=339\text{ MeV}$, $m_{\pi }\left( =m_{\pi }^{\beta }\left( T=0\right) \right)  \simeq138\text{ MeV}$. This set of parameters implies also $m_{0}=5.5$ MeV (the average between the lightest quark masses, $m_{u}\simeq4$ MeV, $m_{d}\simeq7$ MeV), $f_{\pi }=93$ MeV and $G\simeq 5.51$
GeV$^{-2}.$ These values for the parameters are standard in the literature for the case with a 3-dimensional cutoff \cite{Quack}.

At zero temperature we found $a_{0}\simeq0.161$, $a_{2}\simeq -0.043$. These numbers are in agreement with
Weinberg's results \cite{Weinberg} $a_{0}^{W}=7m_{\pi }^{2}/32\pi f_{\pi }^{2}=0.16$ and $
a_{2}^{W}=-2m_{\pi }^{2}/32\pi f_{\pi }^{2}=-0.044$. However, they disagree with the experimental results \cite{Rosselet}, $a_{0}=0.26\pm0.05$ and $a_2=-0.028\pm 0.012$. Nevertheless, our goal here was to find the thermal evolution of the scattering lengths normalized by the zero temperature values.

Our results agree qualitatively well with two previous calculations. The first one \cite{Kaiser}, was done in the frame of Chiral Perturbation Theory at the one loop level, whereas the Linear Sigma Model was used in the second analysis \cite{LM}. In our case, the scattering lengths vary faster as function of temperature compared with the other two papers. However, we disagree qualitatively with the results obtained in \cite{Quack}. This calculation was done also in the frame of the NJL model, but using the imaginary time formalism to compute the thermal corrections.

\section*{ACKNOWLEDGMENTS}

The authors would like to thank financial support from FONDECYT under grants 1051067 and 1060653. M.L. also acknowledges support from the Centro de Estudios Subat\'{o}micos. J.R.A. thanks financial support from the Graduate Program in Physics of the Faculty of Physics, Pontificia Universidad Cat\'olica de Chile.

\end{document}